\documentclass[aps,amsfonts,nofootinbib,superscriptaddress,preprint]{revtex4-1}   
\usepackage{comment}
\usepackage[normalem]{ulem} 
\makeatletter
\usepackage{amsmath}
\usepackage{amssymb}
\usepackage{appendix}
\usepackage{amsbsy}
\usepackage{bm}
\usepackage{epsfig} 
\usepackage{color}
\usepackage{slashed}
\usepackage{soul}
\newcommand{\ee}{\end{equation}}
\newcommand{\bb}{\begin{equation}}
\newcommand{\eqb}{\begin{eqnarray}}

\newcommand{\eqf}{\end{eqnarray}}

\newcommand{\bbM}{\mathbb{M}}
\begin{document}
\title{Gravitons in the Strong-Coupling Regime }
 \author{J.~Gamboa}
\email{jorge.gamboa@usach.cl}
\affiliation{Departamento de  F\'{\i}sica, Universidad de  Santiago de
 Chile, Casilla 307, Santiago, Chile}
\author{R.~MacKenzie}
\email{richard.mackenzie@umontreal.ca}
\affiliation{D{\'e}partement de physique, Universit{\'e} de Montr{\'e}al, Complexe des Sciences,  C.P. 6128, succursale Centre-ville, Montr{\'e}al, QC, Canada, H3C 3J7}
\author{F.~M\'endez}
 \email{fernando.mendez@usach.cl}
\affiliation{Departamento de  F\'{\i}sica, Universidad de  Santiago de
  Chile, Casilla 307, Santiago, Chile}
\begin{abstract}
In the context of gravity in the strong-coupling regime, the propagation amplitude of gravity coupled to relativistic particles undergoing geodesic separation is calculated exactly. Geodesic separation gives rise to boundary terms associated with the $h_\times $ and $ h_+$ graviton components. At low temperatures the propagation amplitude vanishes, implying no graviton propagation in this regime.
\end{abstract}
\maketitle

In a recent article, Parikh, Wilczek and Zahariade \cite{pari} have suggested that the geodesic separation \cite{review} of a pair of freely falling masses includes a stochastic component due to quantum fluctuations of the gravitational field. This noise depends on the state of the gravitational field; surprisingly, for certain types of state (such as thermal or squeezed states), the effect is enhanced and potentially detectable. This gives rise to the tantalizing possibility that the growing array of gravitational wave observatories might also provide the first experimental evidence for quantum gravity effects.

In this letter, we would like to focus on a slight variation of the approach followed in \cite{pari} which suggests new aspects of quantum gravity arising as a consequence of the boundary conditions when the effective action of quantum gravity is calculated via the path integral in Euclidean space.

We begin by considering gravity in the strong-coupling regime 
\cite{isham,pilati1,pilati2,pilati3,isaac2,isaac3,gamboa2}, where
the Poincare group reduces to the Carroll group \cite{levy,levy1,levy2,levy3}. This simplifies the analysis considerably since spatial derivatives can be neglected. The usual Einstein equations $\Box h_{\mu \nu} = -\kappa T_{\mu \nu}$ reduce to
\bb
{\ddot h}_{\mu \nu} = -\kappa T_{\mu \nu}, \label{1}
\ee
to which we must add dynamics for the matter.

Consider a pair of freely falling particles with geodesic separation $\xi_\mu(t)$. Whereas at weak coupling the system would be described by the Einstein-Hilbert action coupled to the two particles (see (2) of \cite{pari}), at strong coupling it is described by the following Lagrangian:
\bb
L= \frac{1}{2} {\dot h_{\mu\nu}}{\dot h^{\mu\nu}} + \frac{m}{2} {\dot \xi_\mu}{\dot \xi^\mu}  + \frac{\alpha}{4} {\dot h}_{\mu \nu} ( {\dot \xi}^{\mu} \xi^{\nu} + \xi^{\mu} {\dot \xi}^{\nu} ), 
\label{3}
\ee
where $m$ is a mass  and 
$\alpha$ is a positive coupling constant
which is proportional to $m$. In what follows, we will set $m=1$ for simplicity.
The Lagrangian \eqref{3} is the starting point of our analysis.

The equations derived from \eqref{3} are
\bb
\begin{array}{rcl}
{\ddot h}_{\mu \nu} &=& -\frac{\alpha}{4}( {\ddot \xi}_{\mu} \xi_\nu  + {\ddot \xi}_{\nu} \xi_\mu  + 2{\dot \xi}_\mu {\dot \xi}_\nu),
\\
{\ddot \xi}_\mu &=&- \frac{\alpha}{2} {\ddot h}_{\mu \nu} \xi^\nu.
\end{array}
 \label{2b}
\ee
Note that $h_{\mu \nu}$ is a cyclic coordinate, so the first of these can be integrated.


The quantum theory is defined via the Euclidean path integral; the propagation amplitude is \cite{hawking1,hawking2}
\bb
G[\phi',\phi; \beta]= \int {\cal D} h {\cal D} \xi \, e^{-S}, \label{action}
\ee
where, of course, we assume some judicious gauge choice has been made and
\[
S=\int_{-\beta/2}^{+\beta/2} dt \, L
\]  
The notation $\phi$ in (\ref{action}) collectively denotes $(h_{\mu \nu}, \xi_\mu)$ so $ G [\phi',\phi; \beta]$ is the amplitude to go from an initial configuration $\phi=(h,\xi)$ at  Euclidean time $-\beta/2$ to a final configuration $\phi'=(h',\xi')$ at Euclidean time $ +\beta/2 $. The path integral is over all configurations interpolating between the initial and final configurations.

In order to compute the effective action of gravity, we integrate over $\xi$. Since the action is quadratic in $\xi$ this gives, up to an overall $h$-independent factor,
\bb
G[h',h; \beta]= \int {\cal D}h ~\det \left[-\delta_{\mu \nu} \partial_t^2 -\frac{\alpha}{2} {\ddot h}_{\mu \nu} \right]^{-1/2}\times ~e^{-\int_{-\beta/2}^{+\beta/2} dt\frac{1}{2} {\dot h}^2},  \label{effec1}
\ee
To simplify the discussion, we define
\bb
\bbM_{\mu \nu} =- \delta_{\mu \nu} \partial_t^2 -\frac{\alpha}{2} {\ddot h}_{\mu \nu}.
 \label{effec2}
\ee

We use the transverse-traceless (TT) gauge in which $h_{\mu \nu}$ is traceless and has the form 
\bb
h_{\mu \nu} = e^+_{\mu \nu} h_+ (t) + e^\times_{\mu \nu}h_\times (t), \label{ss2}
\ee 
where $e^{+,\times}_{\mu \nu}$ are polarization tensors which we assume are constant. If the system describes gravitational waves propagating along the $ {z}$-axis, then the perturbation can be written
\bb
[h] = \begin{pmatrix}
0 & 0& 0& 0  
\\
0 & h_+ & h_\times& 0 
\\
0 & h_\times &-h_+ &0 
\\ 
0 & 0 & 0 & 0 
\end{pmatrix}.
\ee
Ignoring an irrelevant constant, the determinant in \eqref{effec1} is
\bb 
\det \mathbb{M}= \det \begin{pmatrix}
-\partial_t^2 -\frac{\alpha}{2} {\ddot h}_+ & -\frac{\alpha}{2}  {\ddot h}_\times
\\
-\frac{\alpha}{2} {\ddot h}_\times &  -\partial_t^2 +\frac{\alpha}{2} {\ddot h}_+
\end{pmatrix}; \label{det212}
\ee

We choose the polarization $h_\times=0$, giving  
\eqb  
\det \mathbb{M}
&=& \det \left( -\partial_t^2 +\frac{\alpha}{2} {\ddot h}_+\right) \det \left( -\partial_t^2 -\frac{\alpha}{2} {\ddot h}_+\right).
\label{det2121}
\nonumber 
\\
&\equiv&\det \mathbb{M}_1 \det \mathbb{M}_2\eqf 
We focus on the operator $\bbM_1$, which can be factorized as follows
\bb
-\partial_t^2 + \frac{\alpha}{2}{\ddot h}_+  = \left(-\partial_t + {\dot V}\right) \left(\partial_t + {\dot V}\right), \label{mdd1}
\ee 
where  $V(t)$ is defined  through the  Riccati equation 
\bb
-\ddot {V} + {\dot V}^2 = \frac{\alpha}{2}{\ddot h}_+ . 
\label{rica}
\ee

This procedure has become standard in supersymmetric quantum mechanics \cite{witten,cooper} where $V =-\ln \psi_0$ and $\psi_0$ is the ground state (assumed to be of zero energy in SUSY quantum mechanics) of the associated eigenvalue problem to (\ref{mdd1}). 


Using (\ref{mdd1}) we can write the first determinant in (\ref{det212}) as 
\eqb 
\det{\mathbb{M}_1} &=&  \det\left( -\partial_t^2 + \frac{\alpha}{2}{\ddot h}_+\right) = \det \left(-\partial_t + {\dot V}\right) \det \left(\partial_t + {\dot V}\right) \nonumber 
\\ 
&=& \prod_{n=-\infty}^\infty \lambda_n^+\prod_{n=-\infty}^\infty \lambda_n^-
\label{eq-detM1}
\eqf
where $ \{\lambda_n^+, \lambda_n^-\}$ are the eigenvalues of the corresponding operators: 
\begin{subequations}
\eqb 
\left( -\partial_t + {\dot V}\right) \psi^+ &=& \lambda^+\psi^+, \label{s1}
\\
\left( \partial_t + {\dot V}\right) \psi^- &=& \lambda^-\psi^-. \label{s2}
\eqf
\end {subequations}
These equations are easily solved; up to a multiplicative constant, we find
\begin{subequations}
\begin{eqnarray}
\label{solplus}
\psi^{+}(t) &=&e^{+V(t) - \lambda^{+} t},
\\
\label{solminus}
\psi^{-} (t)&=&e^{-V(t) + \lambda^{-} t}.
\end{eqnarray}
\end{subequations}

The eigenvalues are determined by imposing periodic boundary conditions on $\psi^\pm$ on the interval $[-\beta/2,\beta/2]$, giving
\begin{equation}
\label{eq-lambdapm}
\lambda^\pm_n= \frac{\Delta V(\beta)\mp 2n\pi i}{\beta},
\end{equation}
where $\Delta V(\beta) = V(\beta/2) - V(-\beta/2)$ and $n\in\mathbb{Z}$.

The determinant can be evaluated by substituting \eqref{eq-lambdapm} into the logarithm of \eqref{eq-detM1}, giving
\bb
\log(\det\bbM_1) = \sum_{n=-\infty}^{\infty} \log \left[ b^2  + a^2 n^2 \right]
\ee
where to simplify the evaluation of the sum we have written $a=2\pi/\beta$ and $b=\Delta V(\beta)/\beta$. The sum can be evaluated by differentiating with respect to $b$, resulting in an integral representation of $\coth(\pi b/a)$ which can be integrated with respect to $b$, giving
\bb
\det{\mathbb{M}_1} = \sinh^2 \frac{\Delta V(\beta)}{2},
\ee
 where an irrelevant multiplicative constant related to regularization of the summation has been dropped. Our calculation is consistent with others using  $\zeta$-function regularization (for a discussion, see \cite{dunne,esp}).

The calculation of the determinant of the operator $\mathbb{M}_2$  is similar (although associated with another eigenvalue problem) and the result is
\bb
\det{\mathbb{M}_2} = \sinh^2 \frac{\Delta W(\beta)}{2}, 
\ee
where $W$ is a function analogous to $V$ that satisfies a different Riccati equation ({\em cf.\/} \eqref{rica}) 
\begin{equation}
-\ddot {W} + {\dot W}^2 = -\frac{\alpha}{2}{\ddot h}_+ ,
\label{rica2}
\end{equation}
and $\Delta W(\beta) = W(\beta/2) -  W(-\beta/2)$.

The propagation amplitude (\ref{effec1}) becomes 
\eqb
G[h',h; \beta] &=& \int  {\cal D} h_+ \,{\det(\mathbb{M})}^{-1/2} \times ~e^{-\int_{-\beta/2}^{+\beta/2} dt \frac{1}{2} {\dot h_+}^2 } \nonumber 
\\
&=& \int  {\cal D} h_+ 
\frac{e^{-\int_{-\beta/2}^{+\beta/2} dt \frac{1}{2} {\dot h_+}^2 }}{\sinh \left| \Delta V(\beta) \right|
\sinh \left| \Delta W(\beta) \right|}
\label{squirrel}
\eqf

It is interesting to examine the behavior of this expression in the low-temperature limit, that is, when $\beta\to\infty$. The functions $V(t)$ and $W(t)$ depend in a rather complicated manner on $h_+(t)$ through the Riccati equations \eqref{rica} and \eqref{rica2}, respectively. Define $\chi_V$ and $\chi_W$ by
$$
\chi_{V}(t) = e^{-V(t)},
\qquad
\chi_{W}(t) = e^{-W(t)}.
$$
Then \eqref{rica} and \eqref{rica2} become
\begin{equation}
\label{sch1}
\ddot{\chi}_{V} - \frac{\alpha }{2}\,{\ddot h}_+\chi_{V} =0,
\qquad
\ddot{\chi}_{W}  + \frac{\alpha}{2}\,{\ddot h}_+\chi_{W} =0,
\end{equation}
while \eqref{squirrel} reads
\bb
G[h',h; \beta] =
\int  {\cal D} h_+
\frac{e^{-\int_{-\beta/2}^{+\beta/2} dt \frac{1}{2} {\dot h_+}^2 }}
{\sinh \left| \log\frac{\chi_V(+\beta/2)}{\chi_V(-\beta/2)} \right|
\sinh \left| \log\frac{\chi_W(+\beta/2)}{\chi_W(-\beta/2)} \right|}.
\label{hamster}
\ee

It can be shown that under very general conditions one of the factors in the denominator of \eqref{hamster} diverges as $\beta\to\infty$ while the other remains finite.
For example, for the special
class of functions with behavior ${\ddot h}_+\sim t^k$ ($k > 0$), $\chi_{W}$ remains finite while $
\chi_{V} \to \infty$. More generally, $\chi_V$ and $\chi_W$ are zero-energy solutions of Schroedinger-like equations with potentials $\pm \ddot h_+(t)$, respectively. If $\ddot h_+$ is positive as $t\to\pm\infty$, $\chi_V$ diverges exponentially while $\chi_W$ is oscillatory; if $\ddot h_+$ is negative as $t\to\pm\infty$, the roles of $\chi_V$ and $\chi_W$ are reversed; finally, if $\ddot h_+$ changes sign, each of $\chi_V$ and $\chi_W$ will be oscillatory on one side and divergent on the other. In all cases, the denominator in \eqref{hamster} goes to zero in the path integral over a large set of functions ${\ddot h}_+$, and the propagation amplitude goes to zero.

In other words, the graviton does not propagate at low temperatures, at least at strong coupling which is an ingredient for graviton condensation.

As a final comment, we would like to point out that classically the strong-coupling regime, at least mathematically in our context, resembles a Kasner universe for each point of spacetime. The decoupling between neighboring points of spacetime can be recognized as an example of a BKL oscillation \cite{bkl} which is a kind of cosmological singularity.  On the other hand, the vanishing of the propagation amplitude suggests that this cosmological singularity is smoothed out by quantum effects.

This work was supported by Dicyt 041831GR (J.~G.), the Natural Sciences and Engineering Research Council of Canada (R.~M.) and  Dicyt 041931MF (F.~M.). We thank Rafael Benguria for useful discussions.

\end{document}